\documentclass[a4paper]{article}
\usepackage[american]{babel}
\usepackage{fullpage}
\usepackage[latin1]{inputenc}
\usepackage{amsthm,amsfonts,amsmath,url,amssymb}
\usepackage{fullpage}
\usepackage{paralist}
\usepackage{xspace}

\newtheorem{remark}{Remark}
\newcommand{\proba}{\mathit{success}}
\newcommand{\WAS}{W}
\newcommand{\WASmin}{W_{\mathit{min}}}
\newcommand{\Topt}{T_{\mathit{opt}}}

\title{Checkpointing vs. Migration for Post-Petascale Machines}

\author{Franck Cappello \and Henri Casanova \and Yves Robert}

\date{November 2009}

\begin{document}
\maketitle

\bigskip
\begin{center}
\framebox{\textbf{Research Report LIP-2009-32}}

\vspace{1cm}
\centerline{\textbf{\LARGE EXTENDED ABSTRACT}}
\end{center}

\vspace{2cm}

\begin{abstract}
We craft a few scenarios for the execution of sequential and parallel jobs
on future generation machines. Checkpointing or migration, which technique to choose?
\end{abstract}

\bigskip
\section{Introduction}
\label{sec.intro}

From fault-tolerance to resilience~\cite{ijhpca-faulttolerance,ijhpca-exascale-resilience}.
Large machines are subject to failures.
Applications will face resource faults during execution. Fortunately,
failure prediction is there to help. For instance, the system will receive an alarm
when a disk or CPU becomes unusually hot. In that case, the application must
dynamically do something to prepare for, and recover from, the expected failure.
The goal is to compare two well-known strategies:
\begin{itemize}
  \item Checkpointing: purely local, but can be very costly
  \item Migration: requires availability of a spare resource
\end{itemize}

Finally, we assess the cost of periodically checkpoint parallel jobs
in the absence of failure prediction.

\section{Notations}

\begin{itemize}
  \item $C$: checkpoint save time (in minutes)
  \item $R$: checkpoint recovery time (in minutes)
  \item $D$: down/reboot time (in minutes)
  \item $M$: migration time (in minutes)
  \item $N$: total number of cluster nodes
  \item $\mu$: the mean time between failures (e.g., $1/\lambda$ if the failures are exponentially distributed)
\end{itemize}

Obviously, the checkpointing/migration comparison makes sense only if $M < C+D+R$,
otherwise better use the faulty machine as its own spare. Techniques such as
\emph{live migration}~\cite{clark2005lmv} allow for migrating without any disk access,
thereby dramatically reducing migration time.

\section{Sequential jobs}

\subsection{Checkpointing}

We checkpoint just in time before the failure. Each resource is
unavailable during $C+D+R$ time-steps, and this happens every $\mu$
time-steps in average. Hence the global throughput is $$\rho_{cp} =
\frac{\mu}{\mu+C+D+R} \times N$$

\subsection{Migration}

Let us assume we keep $m$ of the $N$ nodes as spares.
We need to ensure that we are never short of a spare machine. We encounter
a problem in the execution if there are more than $m$ resources that
are engaged in migration or rebooting. The probability that, at a given time, a machine
is not migrating or rebooting is:
$$ u = \frac{\mu}{\mu+M+D}\;,$$
and that it is migrating or rebooting is:
$$ v = \frac{M+D}{\mu+M+D}\;.$$

Therefore, the probability that we do not encounter a problem is:
$$
\proba(m) = \sum_{k=0}^{m} \binom{N}{k} u^{N-k} v^{k}\;.
$$

So we need to find the good percentage of spare machines, say $m =
\alpha(\varepsilon) N $, that ``guarantees'' a successful execution with
probability at least $1 - \varepsilon$. Unfortunately, the expression for
$\proba(m)$ doesn't allow for solving the $\proba(m) \geq 1 - \varepsilon$
equation analytically. It must therefore be solved numerically.

Note that $\binom{N}{k} \geq (N/k)^k$. Therefore,
$$\proba(m) \geq \sum_{k=0}^{m} \left( N/k \right)^k u^{N-k}v^{k}\;,$$
which may be a
bit easier to use for numerically solving the equation, and leads to an
overestimation of the number of spares for achieving a probability of
success $1 - \varepsilon$.

\bigskip
Given $m$ spares, the global throughput is $$\rho_m = \frac{\mu}{\mu+M}
\times (N-m)$$

\begin{remark}
When there is a problem with migration, it does mean that the
execution fails, because we cannot find a spare to replace a machine that goes down, and
at that moment, it is too late to checkpoint.
\end{remark}

\section{Parallel jobs}

\subsection{Distribution}
\label{sec.par-job-distrib}

The number of processors required by typical job obeys a strange distribution, which is a two-stage
log-uniform distribution biased to powers of two, see~\cite{feitelson-workload-model}.
We assume something similar but simpler:
\begin{itemize}
  \item let $N=2^{Z}$ for simplicity
  \item the probability that a job is sequential is $\alpha_0 = p_1 \approx 0.25$
  \item otherwise, the job is parallel, and the probability that it uses $2^j$ processors
   is independent of $j$ and equal to $\alpha_j = (1-p_1) \times \frac{1}{Z}$ for $1 \leq j \leq Z = \log_2 N$
\end{itemize}

We assume a steady-state utilization of the whole platform, where all processors are active all the time,
and where the proportion of jobs using any
given number of processors remains constant.
At any time-step, the expectation of the number of jobs that use
$2^j$ processors exactly is $\beta_j$ for $0 \leq j \leq Z$.
The expectation of the total number of jobs running is $K$.
We have:
\begin{eqnarray}
\label{eq.jobs1} K & = & \sum_{j=0}^{Z} \beta_j\\
\label{eq.jobs2} \beta_j & = &\alpha_j K \text{ for } 0 \leq j \leq Z\\
\label{eq.jobs3} N & = & \sum_{j=0}^{Z} 2^j \beta_j
\end{eqnarray}

We derive
$$\frac{N}{K} = \sum_{j=0}^{Z} 2^j \alpha_j = p_1 + \frac{1-p_1}{Z} \sum_{j=1}^{Z} 2^j = p_1 + \frac{1-p_1}{Z} (2N-2)$$
hence the value of $K$, and then that of all the $\beta_j$.

\subsection{Checkpointing}

If a job uses two processors, then the expected interval time between
failures is $\mu/2$. This is because the minimum of two identical
exponential laws is exponential with a doubled parameter.  More generally,
let's call $\mu_k$ the mean of the minimum of $2^k$ i.i.d. variables.
If the variables are exponentially distributed, with scale parameter
$\lambda$, then $\mu_k = 1 / (\lambda 2^k)$.  If the variables are
Weibull, with scale parameter $\lambda$ and shape parameter $a$, then
$\mu_k = \lambda \Gamma(1 + 1/(a 2^k))$.

For $0 \leq k \leq Z$, there are $\beta_k \times 2^k$ processors running jobs with $2^k$ parallel tasks,
hence whose expected interval time between
failures is $\mu_k$. The throughput is given as:
$$\rho_{cp} = \sum_{k=0}^{Z} \beta_k \times 2^k \times \frac{\mu_k}{\mu_k+C+D+R}\;.$$
For the exponential distribution, this becomes:
$$\rho_{cp} = \sum_{k=0}^{Z} \beta_k \times 2^k \times \frac{\frac{1}{\lambda}}{\frac{1}{\lambda}+2^k(C+D+R)}\;.$$

\subsection{Migration}

The probability of running OK is the same as for independent jobs:
$$\proba(m) = \sum_{k=0}^{m} \binom{N}{k} u^{N-k} v^{k}\;.$$
Because there are only $N-m$ machines "really" available, we scale the throughput by
the factor $(N-m)/N)$.
The global throughput now becomes
$$\rho_m = \left( \sum_{k=0}^{Z} \beta_k \times 2^k \times  \frac{\mu}{\mu+2^k M} \right) \times \frac{N-m}{N}$$

\section{Numerical Results}

In this section we present numerical results to understand the impact
of checkpointing vs. migration under a number of scenarios, both in the
"all sequential" case and in the "parallel jobs" case.  All results are
in percentage improvement of migration over checkpointing (negative or
positive values).

All results use the following values:
\begin{itemize}
\item $\mu = $ 1 day, 1 week, 1 month, 1 year;
\item $N = 10,000$, $100,000$, $1,000,000$;
\item $\varepsilon = 10^{-4}$, $10^{-6}$.
\end{itemize}
\noindent
and with particular values of $C=R$, $M$, and $D$ in the following
scenarios.

\subsection{Scenario "today"}

\begin{itemize}
\item $C=R \in [20, 30]$
\item $D \in [1.5, 5]$
\item $M \in [.5, 1.5]$ (32GB on a 10Gbps net)
\end{itemize}

Results in Table~\ref{tab.table_today} for particular values in the above ranges.

\begin{table}
\caption{"Today" scenario: $C=25$, $D=2.5$,
         $M=1$. Percentage improvement of migration over checkpointing. Numbers
         of required spares in parentheses.}
\label{tab.table_today}
\begin{center}
\begin{tabular}{|c|c||c|c||c|c||}
\cline{3-6}
\multicolumn{2}{}{} & \multicolumn{2}{|c||}{Sequential Jobs} & \multicolumn{2}{|c|}{Parallel Jobs} \\
\hline
$\mu$ & $N$  & $\varepsilon =$ $10^4$ & $\varepsilon =$ $10^6$ & $\varepsilon =$ $10^4$ & $\varepsilon =$ $10^6$ \\
\hline
\hline
           & $2^{14}$ & 2.75 (65) & 2.65 (73) & 2081.37 (65) & 2080.30 (73)   \\
1 day    & $2^{17}$ & 2.96 (386) & 2.93 (406) & 2760.06 (386) & 2759.62 (406)   \\
           & $2^{20}$ & 3.03 (2732) & 3.02 (2786) & 3200.37 (2732) & 3200.20 (2786)   \\
\hline
           & $2^{14}$ & 0.31 (16) & 0.27 (20) & 1196.77 (16) & 1196.45 (20)   \\
1 week    & $2^{17}$ & 0.40 (73) & 0.39 (81) & 2158.28 (73) & 2158.15 (81)   \\
           & $2^{20}$ & 0.43 (437) & 0.42 (458) & 2824.48 (437) & 2824.42 (458)   \\
\hline
           & $2^{14}$ & -0.02 (3) & -0.04 (5) & 136.28 (3) & 136.25 (5)   \\
1 month    & $2^{17}$ & 0.00 (8) & 0.00 (10) & 609.60 (8) & 609.59 (10)   \\
           & $2^{20}$ & 0.01 (27) & 0.01 (32) & 1575.36 (27) & 1575.35 (32)   \\
\hline
           & $2^{14}$ & -0.02 (2) & -0.02 (2) & 14.81 (2) & 14.81 (2)   \\
1 year    & $2^{17}$ & -0.00 (3) & -0.00 (4) & 97.57 (3) & 97.57 (4)   \\
           & $2^{20}$ & 0.00 (6) & -0.00 (9) & 471.29 (6) & 471.29 (9)   \\
\hline
\end{tabular}

\end{center}
\end{table}

\subsection{Scenario "2011 HD"}

\begin{itemize}
\item $C=R \in [5, 10]$
\item $D \in [1.5, 5]$
\item $M \in [.5, 1.5]$ (64GB on a 20Gbps net)
\end{itemize}

Results in Table~\ref{tab.table_2011_hd} for particular values in the above ranges.

\begin{table}
\caption{"2011 HD" Scenario: $C=7.5$, $D=2.5$,
         $M=1$. Percentage improvement of migration over checkpointing. Number
         of required spares in parentheses.}
\label{tab.table_2011_hd}
\begin{center}
\begin{tabular}{|c|c||c|c||c|c||}
\cline{3-6}
\multicolumn{2}{}{} & \multicolumn{2}{|c||}{Sequential Jobs} & \multicolumn{2}{|c|}{Parallel Jobs} \\
\hline
$\mu$ & $N$  & $\varepsilon =$ $10^4$ & $\varepsilon =$ $10^6$ & $\varepsilon =$ $10^4$ & $\varepsilon =$ $10^6$ \\
\hline
\hline
           & $2^{14}$ & 0.34 (65) & 0.25 (73) & 773.24 (65) & 772.81 (73)   \\
1 day    & $2^{17}$ & 0.55 (386) & 0.52 (406) & 995.63 (386) & 995.46 (406)   \\
           & $2^{20}$ & 0.62 (2732) & 0.61 (2786) & 1131.29 (2732) & 1131.23 (2786)   \\
\hline
           & $2^{14}$ & -0.03 (16) & -0.08 (20) & 458.73 (16) & 458.59 (20)   \\
1 week    & $2^{17}$ & 0.05 (73) & 0.04 (81) & 796.68 (73) & 796.63 (81)   \\
           & $2^{20}$ & 0.08 (437) & 0.08 (458) & 1012.44 (437) & 1012.42 (458)   \\
\hline
           & $2^{14}$ & -0.03 (3) & -0.06 (5) & 50.04 (3) & 50.02 (5)   \\
1 month    & $2^{17}$ & -0.01 (8) & -0.01 (10) & 236.64 (8) & 236.64 (10)   \\
           & $2^{20}$ & 0.00 (27) & -0.00 (32) & 595.00 (27) & 595.00 (32)   \\
\hline
           & $2^{14}$ & -0.02 (2) & -0.02 (2) & 4.86 (2) & 4.86 (2)   \\
1 year    & $2^{17}$ & -0.00 (3) & -0.01 (4) & 35.06 (3) & 35.06 (4)   \\
           & $2^{20}$ & -0.00 (6) & -0.00 (9) & 182.61 (6) & 182.61 (9)   \\
\hline
\end{tabular}

\end{center}
\end{table}

\subsection{Scenario "2011 SSD"}

\begin{itemize}
\item $C=R \in [4, 6]$
\item $D \in [1.5, 5]$
\item $M \in [.5, 1.5]$ (64GB on a 20Gbps net)
\end{itemize}

Results in Table~\ref{tab.table_2011_ssd} for particular values in the above ranges.

\begin{table}
\caption{"2011 SSD" scenario: $C=5$, $D=2.5$,
         $M=1$. Percentage improvement of migration over checkpointing. Number
         of required spares in parentheses.}
\label{tab.table_2011_ssd}
\begin{center}
\begin{tabular}{|c|c||c|c||c|c||}
\cline{3-6}
\multicolumn{2}{}{} & \multicolumn{2}{|c||}{Sequential Jobs} & \multicolumn{2}{|c|}{Parallel Jobs} \\
\hline
$\mu$ & $N$  & $\varepsilon =$ $10^4$ & $\varepsilon =$ $10^6$ & $\varepsilon =$ $10^4$ & $\varepsilon =$ $10^6$ \\
\hline
\hline
           & $2^{14}$ & -0.00 (65) & -0.10 (73) & 563.73 (65) & 563.40 (73)   \\
1 day    & $2^{17}$ & 0.21 (386) & 0.17 (406) & 719.04 (386) & 718.91 (406)   \\
           & $2^{20}$ & 0.27 (2732) & 0.26 (2786) & 811.69 (2732) & 811.64 (2786)   \\
\hline
           & $2^{14}$ & -0.08 (16) & -0.13 (20) & 337.65 (16) & 337.55 (20)   \\
1 week    & $2^{17}$ & 0.00 (73) & -0.01 (81) & 580.07 (73) & 580.03 (81)   \\
           & $2^{20}$ & 0.03 (437) & 0.03 (458) & 730.30 (437) & 730.28 (458)   \\
\hline
           & $2^{14}$ & -0.03 (3) & -0.06 (5) & 35.92 (3) & 35.90 (5)   \\
1 month    & $2^{17}$ & -0.01 (8) & -0.01 (10) & 174.29 (8) & 174.28 (10)   \\
           & $2^{20}$ & -0.00 (27) & -0.00 (32) & 436.32 (27) & 436.32 (32)   \\
\hline
           & $2^{14}$ & -0.02 (2) & -0.02 (2) & 3.40 (2) & 3.40 (2)   \\
1 year    & $2^{17}$ & -0.00 (3) & -0.01 (4) & 25.00 (3) & 25.00 (4)   \\
           & $2^{20}$ & -0.00 (6) & -0.00 (9) & 134.17 (6) & 134.17 (9)   \\
\hline
\end{tabular}

\end{center}
\end{table}

\subsection{Scenario "2011 Flash"}

\begin{itemize}
\item $C=R \in [1.5, 2]$
\item $D \in [1.5, 5]$
\item $M \in [.5, 1.5]$ (64GB on a 20Gbps net)
\end{itemize}

Results in Table~\ref{tab.table_2011_flash} for particular values in the above ranges.

\begin{table}
\caption{"2011 Flash" scenario:
         $C=1.5$, $D=2.5$, $M=1$. Percentage improvement of migration over
         checkpointing. Number of required spares in parentheses.}

\label{tab.table_2011_flash}
\begin{center}
\begin{tabular}{|c|c||c|c||c|c||}
\cline{3-6}
\multicolumn{2}{}{} & \multicolumn{2}{|c||}{Sequential Jobs} & \multicolumn{2}{|c|}{Parallel Jobs} \\
\hline
$\mu$ & $N$  & $\varepsilon =$ $10^4$ & $\varepsilon =$ $10^6$ & $\varepsilon =$ $10^4$ & $\varepsilon =$ $10^6$ \\
\hline
\hline
           & $2^{14}$ & -0.48 (65) & -0.58 (73) & 245.48 (65) & 245.31 (73)   \\
1 day    & $2^{17}$ & -0.28 (386) & -0.31 (406) & 306.01 (386) & 305.95 (406)   \\
           & $2^{20}$ & -0.21 (2732) & -0.22 (2786) & 339.91 (2732) & 339.89 (2786)   \\
\hline
           & $2^{14}$ & -0.15 (16) & -0.20 (20) & 150.13 (16) & 150.07 (20)   \\
1 week    & $2^{17}$ & -0.07 (73) & -0.08 (81) & 252.08 (73) & 252.06 (81)   \\
           & $2^{20}$ & -0.04 (437) & -0.04 (458) & 310.19 (437) & 310.18 (458)   \\
\hline
           & $2^{14}$ & -0.04 (3) & -0.06 (5) & 14.76 (3) & 14.75 (5)   \\
1 month    & $2^{17}$ & -0.01 (8) & -0.01 (10) & 76.90 (8) & 76.90 (10)   \\
           & $2^{20}$ & -0.00 (27) & -0.00 (32) & 192.75 (27) & 192.75 (32)   \\
\hline
           & $2^{14}$ & -0.02 (2) & -0.02 (2) & 1.33 (2) & 1.33 (2)   \\
1 year    & $2^{17}$ & -0.00 (3) & -0.01 (4) & 10.15 (3) & 10.15 (4)   \\
           & $2^{20}$ & -0.00 (6) & -0.00 (9) & 58.63 (6) & 58.63 (9)   \\
\hline
\end{tabular}

\end{center}
\end{table}

\subsection{Scenario "2011 Flash" + Faster Reboot}

\begin{itemize}
\item $C=R \in [1.5, 2]$
\item $D \in [0, 0.5]$
\item $M \in [.5 1.5]$ (64GB on a 20Gbps net)
\end{itemize}

Results in Table~\ref{tab.table_2011_flash2} for particular values in the above ranges.

\begin{table}
\caption{"2011 Flash + Faster Reboot" scenario: $C=1.5$, $D=0.25$,
         $M=1$. Percentage improvement of migration over checkpointing. Number
         of required spares in parentheses.}
\label{tab.table_2011_flash2}
\begin{center}
\begin{tabular}{|c|c||c|c||c|c||}
\cline{3-6}
\multicolumn{2}{}{} & \multicolumn{2}{|c||}{Sequential Jobs} & \multicolumn{2}{|c|}{Parallel Jobs} \\
\hline
$\mu$ & $N$  & $\varepsilon =$ $10^4$ & $\varepsilon =$ $10^6$ & $\varepsilon =$ $10^4$ & $\varepsilon =$ $10^6$ \\
\hline
\hline
           & $2^{14}$ & -0.21 (30) & -0.27 (35) & 131.39 (30) & 131.32 (35)   \\
1 day    & $2^{17}$ & -0.08 (155) & -0.10 (168) & 161.38 (155) & 161.35 (168)   \\
           & $2^{20}$ & -0.04 (1024) & -0.05 (1056) & 177.39 (1024) & 177.38 (1056)   \\
\hline
           & $2^{14}$ & -0.09 (9) & -0.12 (12) & 80.95 (9) & 80.92 (12)   \\
1 week    & $2^{17}$ & -0.03 (33) & -0.04 (39) & 134.46 (33) & 134.45 (39)   \\
           & $2^{20}$ & -0.01 (174) & -0.01 (188) & 163.10 (174) & 163.10 (188)   \\
\hline
           & $2^{14}$ & -0.02 (2) & -0.04 (3) & 7.52 (2) & 7.51 (3)   \\
1 month    & $2^{17}$ & -0.01 (5) & -0.01 (7) & 40.93 (5) & 40.93 (7)   \\
           & $2^{20}$ & -0.00 (14) & -0.00 (17) & 103.70 (14) & 103.70 (17)   \\
\hline
           & $2^{14}$ & -0.01 (1) & -0.02 (2) & 0.67 (1) & 0.66 (2)   \\
1 year    & $2^{17}$ & -0.00 (2) & -0.00 (3) & 5.14 (2) & 5.14 (3)   \\
           & $2^{20}$ & -0.00 (4) & -0.00 (6) & 30.95 (4) & 30.95 (6)   \\
\hline
\end{tabular}

\end{center}
\end{table}

\subsection{Scenario "2015"}

\begin{itemize}
\item $C=R \in [0, .15]$
\item $D \in [0, .5]$
\item $M \in [.5, 1.5]$ (128GB on a 40Gbps net)
\end{itemize}

Results in Table~\ref{tab.table_2015} for particular values in the above ranges.

\begin{table}
\caption{"2015" scenario: $C=0.05$, $D=0.25$, $M=1$. Percentage
         improvement of migration over checkpointing. Number of required spares
         in parentheses.}
\label{tab.table_2015}
\begin{center}
\begin{tabular}{|c|c||c|c||c|c||}
\cline{3-6}
\multicolumn{2}{}{} & \multicolumn{2}{|c||}{Sequential Jobs} & \multicolumn{2}{|c|}{Parallel Jobs} \\
\hline
$\mu$ & $N$  & $\varepsilon =$ $10^4$ & $\varepsilon =$ $10^6$ & $\varepsilon =$ $10^4$ & $\varepsilon =$ $10^6$ \\
\hline
\hline
           & $2^{14}$ & -0.41 (30) & -0.47 (35) & -47.52 (30) & -47.54 (35)   \\
1 day    & $2^{17}$ & -0.28 (155) & -0.30 (168) & -55.58 (155) & -55.58 (168)   \\
           & $2^{20}$ & -0.24 (1024) & -0.25 (1056) & -58.81 (1024) & -58.81 (1056)   \\
\hline
           & $2^{14}$ & -0.12 (9) & -0.15 (12) & -28.92 (9) & -28.94 (12)   \\
1 week    & $2^{17}$ & -0.06 (33) & -0.07 (39) & -48.25 (33) & -48.25 (39)   \\
           & $2^{20}$ & -0.04 (174) & -0.04 (188) & -55.84 (174) & -55.84 (188)   \\
\hline
           & $2^{14}$ & -0.02 (2) & -0.04 (3) & -2.25 (2) & -2.25 (3)   \\
1 month    & $2^{17}$ & -0.01 (5) & -0.01 (7) & -13.47 (5) & -13.48 (7)   \\
           & $2^{20}$ & -0.00 (14) & -0.00 (17) & -37.62 (14) & -37.62 (17)   \\
\hline
           & $2^{14}$ & -0.01 (1) & -0.02 (2) & -0.20 (1) & -0.21 (2)   \\
1 year    & $2^{17}$ & -0.00 (2) & -0.00 (3) & -1.52 (2) & -1.52 (3)   \\
           & $2^{20}$ & -0.00 (4) & -0.00 (6) & -9.91 (4) & -9.91 (6)   \\
\hline
\end{tabular}

\end{center}
\end{table}

\subsection{Summary}

\begin{itemize}
\item Sequential jobs: forget migration
\item Parallel jobs: prefer migration,
until checkpointing costs dramatically reduce
(in proportion of migration costs)
\end{itemize}

\section{Impact of failure prediction}

In this section we deal with the case where no failure prediction is available.
The idea is to checkpoint periodically. This raises two questions:
\begin{enumerate}
  \item How to determine the optimal period?
  \item What is the impact on platform throughput?
\end{enumerate}

Question~1 has received some attention in the literature for uni-processor jobs.
Let $T$ be the period, i.e. the time between two checkpoints, let $C$ be the checkpoint duration time,
and $\mu$ the expected interval time between failures. We compute $\WAS$, the expected percentage
of time lost, or ``wasted'', as in~\cite{wingstrom-phd}:
\begin{equation}
\label{eq.young}
\WAS = \frac{C}{T} + \frac{T}{2\mu}
\end{equation}

The first term in the right-hand side of Equation~\ref{eq.young}
is by definition, because there are $C$ time-steps devoted to checkpointing every $T$ time-steps.
The second term accounts for the loss due to failures
and is explained as follows: every $\mu$ time-steps, a failure occurs, and we lose
an average of $T/2$ time-steps. Note that because the checkpoint and failure rates are independent,
the quantity $T/2$ does not depend upon the failure distribution (Poisson, Weibull, etc).
$\WAS$ is minimized for $\Topt = \sqrt{2 C \mu}$. This is Young's approximation~\cite{young74}.
The corresponding minimum waste is $\WASmin = \sqrt{\frac{2C}{\mu}}$.

\bigskip
Equation~\ref{eq.young} does not account for recovery time $R$ after each failure. A more accurate
expression is the following:
\begin{equation}
\label{eq.bob-riton}
\WAS = \frac{C}{T} + \frac{\frac{T}{2}+R+D}{\mu}
\end{equation}
Now in the right-hand side we state that every $\mu$ time-steps, a failure occurs, and we lose
an average of $\frac{T}{2}+R+D$ time-steps. $\WAS$ is minimized for the same value $\Topt = \sqrt{2 C \mu}$
as before, but the
corresponding minimum waste becomes
\begin{equation}
\label{eq.bob-riton2}
\WASmin = \frac{R+D}{\mu} + \sqrt{2\frac{C}{\mu}}
\end{equation}

Note that this is different from the first-order approximation given by Daly~\cite[equations (10) and (12)]{daly04}
because we target the steady-state operation of the platform rather than the optimization of the expected
duration of a given job.

It turns out that $\WASmin$ may become larger than $1$ when $\mu$ gets very small, a situation
which is more likely to happen with jobs requiring many processors. In that case the application is not progressing any more.
To solve for $\WASmin \leq 1$ in
Equation~\ref{eq.bob-riton2}, we let $\nu = \frac{1}{\sqrt{\mu}}$ and derive
$$\nu^2 (R+D) + \nu \sqrt{2C} -1 \leq 0$$
We get $\WASmin \leq 1$ if $\nu \leq \nu_b$ (hence $\mu \geq 1/\nu_b^2$) with
$$\nu_b = \frac{-\sqrt{2C}+\sqrt{2C+4(R+D)}}{2(R+D)}.$$
In all cases, the minimum waste is
$$\min(\WASmin,1)$$

\subsection{Independent jobs}

We simply write that the throughput is
$$\rho = (1 - \WASmin) N $$

\subsection{Parallel jobs}

We assume the same distribution of parallel jobs as in Section~\ref{sec.par-job-distrib},
and we keep the same notations $K$ (number of jobs), $\beta_k$ for $1 \leq k \leq Z = \log_2 N$
(number of jobs of size $2^k$), and $\mu_k$ (expected interval time between failures
for a job using $2^k$ processors).

With $2^k$ processors we use $\mu_k$ instead of $\mu$ in Equation~\ref{eq.bob-riton}
to derive the minimum waste $\WASmin(k)$.
The throughput becomes
$$\rho = \sum_{k=0}^{Z}  (1 - \WASmin(k)) 2^k \beta_k  $$

\subsection{Numerical Results}

Here is a typical result for parallel jobs:
\begin{itemize}
\item $C= D = R = 1$
\item $\mu = $ 1 month or 1 year
\item $p_1 = 0.25$
\end{itemize}

Results in Table~\ref{tab.yield} for particular values of $N$.

\begin{table}
\caption{Yield $\rho/N$ for $C=D+R=1$ and $p_1 = 0.25$. Parallel jobs with $p_1 = 0.25$.}
\label{tab.yield}
\begin{center}

\begin{tabular}{|c|c|c|}
\hline
$N$  & Yield ($\mu = $ 1 month) & Yield ($\mu = $ 1 year)\\
\hline
\hline
$2^{8}$ & 90.8\%  & 97.5\% \\
$2^{11}$ & 69.9\% & 92.6\%  \\
$2^{14}$ & 13.5\% & 76.3\%  \\
$2^{17}$ & 01.7\% & 22.1\%\\
$2^{20}$ & 00.2\% & 02.8\% \\
\hline
\end{tabular}

\end{center}
\end{table}

\section{Conclusion}
\label{sec.conclusion}

New software/hardware techniques are needed in order
to reduce checkpoint, recovery, and migration times. This is a condition
for parallel jobs to execute at a satisfying rate
on future massively parallel machines.

As for migration, we point out another requirement, namely being able to rely on accurate failure predictions.

Another direction is to design "self-fault-tolerant" algorithms (e.g. asynchronous iterative algorithms)
whose execution can progress in the presence of local faults.
Also, replication techniques should be investigated: despite the resource costs induced by duplicating the same tasks on different processors,
replication can dramatically increase the reliability of the whole application.

Most likely, parallel jobs will be deployed on large-scale machines through a mix of all previous techniques
(checkpointing, migration, replication, self-tolerant variants).

\bibliographystyle{abbrv}
\bibliography{biblio}

\end{document}